# The Initial Calibration Date of the Antikythera Mechanism after the Saros spiral mechanical Apokatastasis

**A.Voulgaris[1], C.Mouratidis[2], A.Vossinakis[3]**

[1]City of Thessaloniki, Directorate Culture and Tourism, Thessaloniki, GR-54625, Greece,

[2]Merchant Marine Academy of Syros, GR-84100, Greece,

[3]Thessaloniki Astronomy Club, Thessaloniki, GR-54646, Greece

[1]Corresponding author arisvoulgaris@gmail.com

**Abstract**

*This work analyzes the phase correlation of the three lunar cycles and the Saros/Exeligmos cycle, after the study of the chapter "About Exeligmos" in "Introduction to the Phenomena" by Geminus. Geminus refers that each Exeligmos Cycle began on very specific and rare dates, when the Moon positioned at the points of the three lunar cycles beginning: New moon at Apogee and at the Node. The extremely large duration of the Annular Solar eclipse occurred on December 22$^{th}$ 178BC (Saros series 58) marks the start of the "Prominent Saros Cycle Apokatastasis". In the next day, December 23$^{rd}$ 178BC, the Winter Solstice started. During these two neighboring dates, the celebration of the religious festival of Isia started in Egypt and the Hellenistic Greece. After the analysis of the Mechanism's Parapegma events specific position, December 22$^{nd}$/23$^{rd}$ 178BC is an ideal, functional and representative initial date, in order to calibrate the initial position of the Mechanism's pointers.*

## 1.1 Introduction

The Antikythera Mechanism, was a sophisticated geared machine of the Hellenistic era which could calculate and depict the position of the Sun in the Zodiac sky, the lunar phases, to predict the athletic games starting date and one of the most important astronomical events: the solar and the lunar eclipses (Freeth et al. 2006; Freeth et al. 2008).

The eclipse events were engraved on the Saros spiral located on the Back plate, which was divided in 223 subdivisions/cells. Each cell corresponded to a synodic lunar month. By rotating the Lunar Disc-input of the Mechanism (Voulgaris et al. 2018b), the Saros pointer slowly rotated, and during its rotation transited the cells, on some of which there were some glyphs symbols, which were the eclipse event information (Freeth et al. 2006, Freeth 2014, Anastasiou 2016a).

The Saros cycle (Ptolemy in Book IV2 in Almagest, refers the word *Περιοδικός-Periodicos*, Ptolemy 1898; Toomer 1984) was a period of 223 synodic months, equal to 18 years 11 days and 8 hours. After this time span, the sequence of the eclipse events repeated, by (about) an 8 hour delay. Therefore, an eclipse with similar characteristics occurs about 120° further to the west on the sky.

After a period of three Saros cycles - named Exeligmos - which is equal to 54 years and 34 full days, the eclipse event occurs at about the same place. Through time, the solar eclipse path of the successive Saros/Exeligmos slightly changes position towards to the North or to the South Pole. The Saros/Exeligmos series starts and ends with a partial solar eclipse, visible only from one of earth's poles.

## 1.2 Queries regarding the starting position of the Saros and Metonic pointers

In order to use a measuring instrument, a *reference point* is needed, before the measuring procedure. The definition of the *reference point*/starting position, is called calibration of the instrument. The calibration is a procedure in order to set a measuring machine on its characteristic "*zero level*"/initial position. After this, any measurement relative to the "*zero level*", can be achieved. The "*zero level*"- *reference point* is a point/level which presents a rare specificity, having unique characteristic(s), in order to be clearly detected and easily recognized in relation to the rest points, so that it can be easily selected in every calibration of the machine. For this reason, a specific *zero point* position could be the lowest measurable limit (e.g. the temperature of the absolute zero kelvin), the highest limit (e.g. the Summer solstice is defined, when the sun is located on its highest altitude) or a rare coincidence (e.g. 0° Celsius is the unique temperature in which the water coexists in a solid, liquid and gas condition). In contrast, all of the rest points above, below or in between the specific reference point(s), are less important and difficult to be recognized, because none of them is unique. Moreover, the reference point is selected with the minimum possible parameters and conditions, in order for the calibration procedure to be the same every time. For example, for the calibration of a spectrum captured by a spectrograph, a spectral lamp of an element radiating emission lines with characteristic constant wavelength is used. During the calibration, the selected spectral lamp is stationary. Of course, for this specific calibration, someone could use a fast moving spectral lamp with an arbitrary velocity of "*165km/sec*", but this condition increases the difficulty of the precise calibration every time, because the necessary parameters (precise repetition of the specific velocity) and the errors (wavelength shifting/Doppler-Fizeau effect), increase. Therefore, the use of a non-moving spectral lamp is easier and more preferable, creating the minimum equal/same conditions each time the spectrograph is calibrated and the errors are minimized or do not exist.

The special design, the very large number of parts and the complex construction of the Antikythera Mechanism lead to the conclusion that it was used in order to measure/calculate the time presenting the exported results/calculations via its pointers and scales. It is obvious that the manufacturer of the Mechanism designed/constructed his creation and engraved the specific Saros eclipse sequence events, for a specific Epoch/initial date (Freeth 2014; Carman and Evans 2014; Jones 2017; Iversen and Jones 2019; Freeth 2019; Jones 2020 – based on recent observations by the Authors, a new starting date is presented in this paper). Therefore, the Mechanism pointers require an initial positional/date calibration before the start of the measuring procedure. The Metonic and the Saros spirals have the same measuring unit/cell, which is the lunar synodic month. Detecting the start of both spirals leads to the calculation of the pointers position and therefore after starting the Mechanism operation, all the pointers exhibit correct results every time.

A question arises, regarding the starting position of the Metonic and Saros pointers:
- *When the Metonic pointer was positioned on the spiral start on cell-1, in which synodic cell was the Saros pointer aimed?*
- *Could it be any of 223 cells, randomly selected by the ancient manufacturer?*
If the answer is "*yes, it could be in any of the Saros cells, e.g. the Saros cell-219*", then it is hard to explain why the ancient manufacturer did not change the Saros eclipse events sequence, so that the "*Saros cell-219*" cell to be *cell-1*/start of Saros spiral (and the rest of the events to follow this cell), so that both the Metonic and Saros pointers would start from their corresponding *cell-1*, avoiding the repositioning of the Saros pointer after four cells. This very simple change could be easily achieved, without any problem on the sequence or any other mechanical malfunction.
The possibility that the Saros spiral pointer was randomly selected during the start of the Metonic spiral/*cell-1* at the initial calibration date, adds complexity, lack of symmetry, displays a "bad mechanical aesthetic", makes the calibration of the Mechanism more difficult and creates an unjustified, "non-logic" philosophy of the design. Additionally, if the number of the ("random") starting cell is forgotten or lost, this would be a "total disaster" for the calibration, making the Mechanism non-reliable. At the same time it would be extremely difficult to find again the "random" starting cell, by checking the Mechanism results during its operation.
Therefore, it is difficult to justify the possibility of the "random selected" position of the Saros pointer, when the Metonic pointer is located on *cell-1*.

An ideal and functional initial calibration date of the Antikythera Mechanism dials, would be a date on which the Saros and Metonic pointers are as close as possible to their corresponding spiral beginnings.

## Theory

## 2.1 The Apokatastasis of the lunar cycles on Geminus' work

*Geminus of Rhodes* (Geminus 1880/2002; Evans and Berggren 2006), the ancient astronomer of era 50BC, in his work "*Εἰσαγωγή εἰς τα Φαινόμενα*" (*Introduction to the Phenomena*) gives a lot of information regarding astronomy, the measuring of time and the calendar of that era. He describes in detail the sun and moon positions and their cycles. On the chapter XVIII "*Περί Εξελιγμού*" (*About Exeligmos*) he refers to very useful information regarding the lunar cycles and the eclipses: "*Exeligmos is the shortest period which includes an integer number of synodic lunar months, an integer number of days and an integer number of lunar returns"*. The ancient Greek phrase used for "*lunar returns*" by Geminus is "*Ἀποκαταστάσεις τῆς Σελήνης*". The word "Ἀποκατάστασις" (*Apokatastasis*, in singular) means *return to the (same) starting position/place, as it was on its beginning*, i.e. a reset position. Ptolemy 1898 in Almagest, defines the Apokatastasis (Chap. 3a, page 192, 16) and extensively refers the phrases "*Ἀποκαταστάσεις ἀνωμαλίας*", "*Ἀποκαταστάσεις πλάτους*", "*Ἀποκαταστατικός χρόνος*" (corresp. "*returns in anomaly*", "*returns in latitude*", Toomer 1984 Book IV2, "time span between two successive resets to the starting position"). Aristotle, Asclepius, Epicurus, Diodorus Ciculus, Sextus Empiricus, Censorinus etc. refer this word on their works (Theodorou 1958). The word "*Apokatastasis*" is also referred many times on the Front cover inscription of the Antikythera Mechanism (Anastasiou et al. 2016b). In Greek language the word "*Apokatastasis*" is also used on Archaeology (e.g. apokatastasis of monuments-restoration) and on the Orthopedics Surgery (e.g. apokatastasis of bones).

Geminus also refers: "…ὁ δε χρόνος ὁ ἀπό τῆς ἐλάχιστης κινήσεως ἐπί την ἐλαχίστην κίνησιν, Ἀποκατάστασις καλεῖται", i.e. the time span between the minimum (lunar) motion (i.e. the angular velocity of the lunar motion) to the next minimum, is called *Apokatastasis* (of the lunar anomaly period-return to the start of the Anomalistic month). He also states that the lunar minimum angular velocity is 11° 06' 35" per day (when the moon is at *Apogee*) and half an Anomalistic month later (when the moon is at *Perigee*), the maximum angular velocity is 15° 14' 35" per day. Therefore, the *Apokatastasis* of the Anomalistic cycle is measured from a specific starting position (and not by a random position), which is the lunar *Apogee* and it is the time the moon takes to return to *Apogee* (from one *Apogee* to the next one) Fig. 1, 2.

Based on Geminus reference, the *Apokatastasis* of a lunar cycle is defined by two mandatory parameters:
1) The specific defined starting position of the cycle i.e. the reference point, and
2) The time duration to complete a full cycle, so as to return to the initial reference point.

The first mandatory parameter is necessary in order to begin the measurement of time. For example, the ideal (positional) *reference point* for the sun's altitude measurement is the point where the sun crosses the local meridian. On this point, the sun's daily maximum altitude is measured. The *reference point* for the measurement of the solar tropical year *Apokatastasis* is the Vernal Equinox, a very specific and important point of the sun's position (and not a randomly selected point/day out of the 365 days). These *reference points* were extensively used by astronomers, mathematicians and geometers of the ancient era and are also used even today, because the selected points act as a "*compass*" for calculations in geometry, in space and on time.

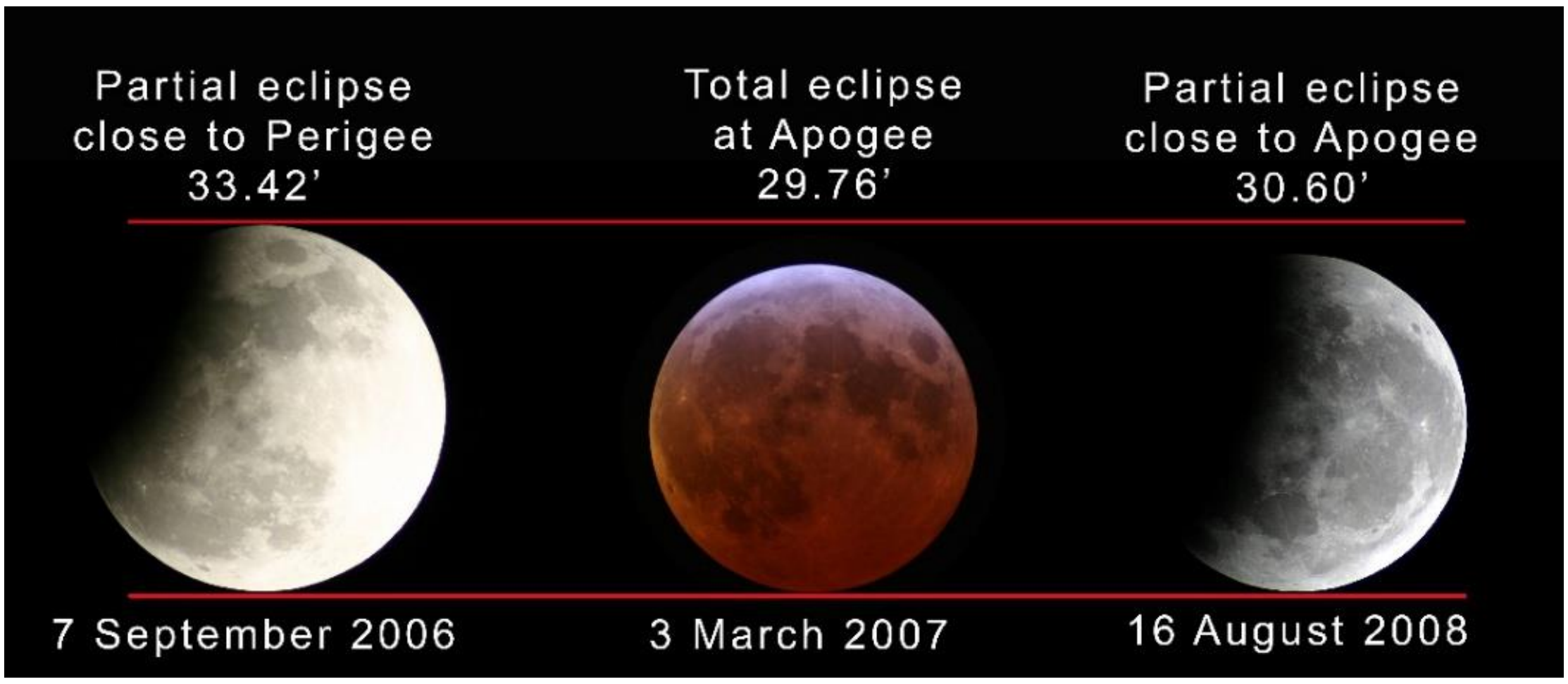

**Fig. 1** Three Lunar eclipses with the moon located in different distances from the Earth. Center: the total lunar eclipse of March 3rd 2007 occurred very close to the Apokatastasis of the Anomalistic lunar cycle-Apogee. One Sar before and after this event a total solar eclipse close to Perigee, with a relative long duration of 04m 09s occurred. Left: the partial lunar eclipse of September 7th 2006 occurred half an Anomalistic month after its Apokatastasis, i.e. very close to Perigee. Right: the partial lunar eclipse of August 16th 2008 occurred when the moon was at an intermediate distance from the Earth. For capturing these images, an 150/1200mm achromatic refractor, was used. All images on same scale, by the first author. On-line color image.

The time span of a lunar Anomalistic cycle starting by any randomly selected phase e.g. $0.732\pi$ to $(0.732\pi+2\pi)$, could be considered as a duration of an Anomalistic month, but it is not an *Apokatastasis* of the Anomalistic cycle, because the *Apokatastasis* is measured from its starting position/reference point, i.e. when the moon is located at *Apogee*. Therefore, the duration of a lunar cycle and the *Apokatastasis* of the same lunar cycle are not the same.

For each lunar cycle Apokatastasis, the ancient astronomers adopted a unique and specific starting point for good reason: the projected lunar motion in the sky is difficult to calculate with accuracy and seems "strange". Therefore it was difficult to predict the lunar position or angular velocity by choosing a random starting position. Moreover, all of the lunar velocity and position values relative to the ecliptic appear twice on each cycle (except the min and max values). Additionally, by adopting a specific reference point for the Apokatastasis was useful to avoid astronomical observations with different and random calibration points, making difficult their management.

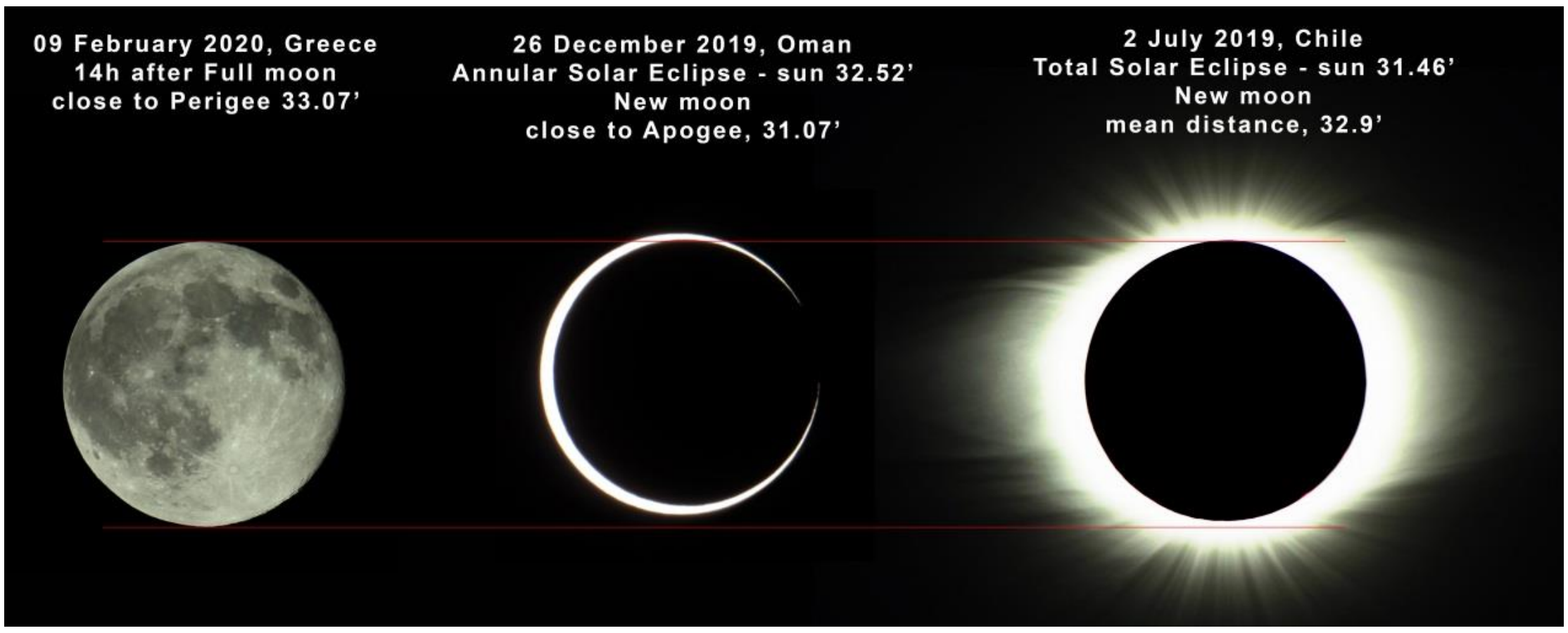

**Fig. 2**, Left: the angular diameters of the Full moon at Perigee on February 9th 2020, relative to the New moon close to Apogee, during the Annular solar eclipse of 26th December 2019 (center) (http://xjubier.free.fr/en/site_pages/solar_eclipses/TSE_20190702_pg01.html), observed from Sarqiya Dessert, Oman. The magnitude of this eclipse was 0.959 and the central duration 2m 54s. Note that the solar angular diameter is also smaller, than the diameter of the Full moon. The red lines are the limits of the Full moon diameter (the upper edge of the Moon is in contact to the top red line in all photos). Right: The Lunar disc (New moon) covers the solar disc during the total solar eclipse of July 2nd 2019 (http://xjubier.free.fr/en/site_pages/solar_eclipses/ASE_20191226_pg01.html), observed from Cerro Tololo Inter-American Observatory, near the Atacama Desert, in north Chile (http://www.ctio.noao.edu/noao/node/14748). The solar corona during totality, is visible with the naked eye. The east/west coronal streamers and the polar plumes, are visible. The magnitude of this eclipse was 1.036 and the totality duration 2m 01s. For capturing the images, an apochromatic refractor 65/450mm, was used. All images on same scale, captured/processed by the authors. On-line color image.

In the same manner, the *Apokatastasis* of the synodic cycle/month, is the return of the moon on its starting position i.e. in syzygy to the sun (New moon) and also *Apokatastasis* of the Draconic cycle, is the return of the moon to the same Node.

Applying the same definition for Exeligmos to the Saros cycle (*Periodicos*), Saros is the shortest period consisting by integer numbers of the lunar cycles *Apokatastasis*, but not an integer number of days (because of the 8 hours delay to the next cycle) i.e. 223 synodic lunar months, 242 draconic *Apokatastasis* and 239 anomalistic *Apokatastasis* (equal to 6585.32$^d$). This means that the synodic, the anomalistic and the draconic cycles return to their corresponding starting positions: the New moon returns at *Apogee* and at the same time it also returns to the same Node.

This very important argument defines that the Saros and the Exeligmos cycle *Apokatastasis* start at very specific and very rare dates, in which all of the lunar cycles are in their *Apokatastasis*/reset position (see Chap. 4).

Finally, adopting an integer number of lunar cycles, guarantees that the lunar motions are repeated on (about) the same relative position, thus offering an eclipse with (about) the same geometrical characteristics, eclipse type and duration. The Exeligmos cycle additionally guarantees that this eclipse will occur at (about) the same time, when observed from the same place (Oppolzer 1962; Neugebauer 1975).

Because the numbers 223 and 239 are prime numbers, Saros is the minimum period of the resonance of the three cycles. A half Saros ($9^y$ and about $5.65^d$), named Sar period, presents a $\pi$-phase resonance on two of the three cycles i.e. 111.5 synodic months equals to 119.5 anomalistic months and also equals to 121 draconic cycles. So, after a Sar period, the eclipses are repeated in their "inversed" geometrical characteristics e.g. a total solar eclipse with the moon located at *Perigee* (very long eclipse duration), after one Sar period corresponds to a total lunar eclipse at lunar Apogee (long eclipse duration) and at the same angular distance from the Node. Therefore, when the solar eclipse is visible from the northern hemisphere, after one Sar period the moon passes from the northern part of the Earth's shadow.
The Sidereal lunar cycle does not correlate to an integer number with Exeligmos/Saros: Geminus refers that one Exeligmos is equal to 723 sidereal cycles+32°. This means that after one Saros cycle, the moon is located at about 10.7° (varies around this number) eastern to its initial position in the sky i.e. the sky background changes on each Saros repetition event (during a total solar eclipse, the brightest stars are visible by naked eye). After one Exeligmos cycle, the eclipse events will occur on the next zodiac constellation.

## Calculation

### 3.1 Data mining by the analysis of the lunar cycles Apokatastasis -The *"Prominent Saros Apokatastasis Cycle"*

Chapter 3 analysis indicates that both the Exeligmos and the Saros Cycles, started when the synodic, anomalistic and draconic cycles were located on their *Apokatastasis* position i.e.
1) New moon phase,
2) New moon is located at *Apogee* and
3) New moon is located on a Node.
This observation leads to a number of very important conclusions, regarding the starting date of a Saros/Exeligmos Cycle.
Because the New Moon is located on a Node, this means that during the start of the Saros Cycle, a solar eclipse will occur. At the same time, the New moon is located at *Apogee* position and this means that the solar eclipse will certainly be an annular eclipse.
Therefore, the Saros/Exeligmos Cycles start with an annular solar eclipse. Because the New moon during the start of the Saros/Exeligmos is exactly located at *Apogee* position i.e. the greatest distance from the Earth, the lunar angular diameter is the minimum possible and at the same time the moon has its lowest angular velocity. A moon with a small angular diameter, covers the minimum solar disc area (small ratio lunar/solar diameter-eclipse magnitude) and transits the solar disc with its lowest possible velocity. This means that the central duration of this specific annular solar eclipse is extremely large (the central duration is the duration when the lunar disc is fully projected inside the solar disc i.e. between 2nd and 3rd contacts). The combination of the lowest lunar angular diameter and the slowest lunar angular velocity, increases dramatically the central duration of the eclipse. This particular annular eclipse is usually called “a *very deep annular eclipse*”.

The position of the Earth on *Perihelion* (*Periapsis*) or *Aphelion* (*Apoapsis*), does not affect significantly the duration of the eclipse:

The angular diameter of the moon varies between 29.30 arc min (*Apogee*) – 34.10 arc min (*Perigee*) and the solar angular diameter varies between 31.46 arc min (*Aphelion*) – 32.53 arc min (*Perihelion*). The maximum value of an eclipse magnitude, when the moon is located at *Apogee* and the Earth at *Aphelion* is $Moon_{apogee}/Sun_{aphelion}$ = 0.931341 and the minimum value, when the moon is located at *Apogee* and the Earth at *Perihelion* is $Moon_{apogee}/Sun_{perihelion}$ = 0.90070, a difference of 0.03063. For an annular eclipse with central duration of 10 min, this difference translates to a shorter central duration by about 18.3 sec. So, the main/strong determinant factor for the type and the duration of a solar eclipse is the Earth-Moon distance (Apogee) for Annular eclipses (see Table 1) and for Total solar eclipses is the Moon Perigee. The duration of an eclipse also depends on the observing place (as also Geminus refers) and the gamma of an eclipse (https://eclipse.gsfc.nasa.gov/SEhelp/SEglossary.html).

**Table 1** The longest central duration Annular eclipses between 340 BC (-339) and 87BC (-86). 15 out of 31 solar eclipses belong to the Saros series 58 and present an extremely long central duration. On each successive eclipse of Saros series 58, the central duration increases significantly, approaching the longest central duration of the eclipse of 178BC and after this, it gradually decreases. The Gamma value of Saros series 58 increases through time which means that the shadow path of the successive eclipses is moving closer to the North Pole (see NASA eclipse catalogue). Note that in the years after 300BC, the Annular eclipse of December 22nd 178BC is the only one that takes place very close to one of the four tropical points

| Calendar Date | Saros series | Eclipse Magnitude of the Annular eclipse | Central duration |
|---|---|---|---|
| 340 BC, Sep 15 | 58 | 0.9293 | 07m41s |
| 336 BC, Jan 08 | 55 | 0.9200 | 09m16s |
| 336 BC, Dec 28 | 65 | 0.9438 | 07m28s |
| 322 BC, Sep 26 | 58 | 0.9256 | 08m14s |
| 318 BC, Jan 19 | 55 | 0.9225 | 08m47s |
| 304 BC, Oct 07 | 58 | 0.9223 | 08m49s |
| 300 BC, Jan 30 | 55 | 0.9254 | 08m15s |
| 286 BC, Oct 18 | 58 | 0.9195 | 09m27s |
| 282 BC, Feb 10 | 55 | 0.9286 | 07m43s |
| 268 BC, Oct 28 | 58 | 0.9174 | 10m06s |
| 264 BC, Feb 20 | 55 | 0.9321 | 07m12s |
| 253 BC, Jul 16 | 61 | 0.9483 | 07m13s |
| 250 BC, Nov 09 | 58 | 0.9158 | 10m44s |
| 235 BC, Jul 27 | 61 | 0.9445 | 07m46s |
| 232 BC, Nov 19 | 58 | 0.9150 | 11m19s |
| 214 BC Nov 30 | 58 | 0.9148 | 11m47s |
| 199 BC, Aug 18 | 61 | 0.9363 | 08m28s |
| 196 BC, Dec 11 | 58 | 0.9153 | 12m04s |
| 181 BC, Aug 28 | 61 | 0.9323 | 08m37s |
| **178 BC, Dec 22** | **58** | **0.9165** | **12m08s** |
| 159 BC, Jan 01 | 58 | 0.9184 | 11m54s |
| 141 BC, Jan 13 | 58 | 0.9208 | 11m25s |
| 132 BC, Feb 01 | 77 | 0.9121 | 07m45s |
| 123 BC, Jan 23 | 58 | 0.9237 | 10m42s |
| 118 BC, Oct 21 | 80 | 0.9153 | 07m08s |
| 114 BC, Feb 12 | 77 | 0.9179 | 07m43s |
| 105 BC, Feb 03 | 58 | 0.9270 | 09m50s |
| 100 BC, Oct 31 | 80 | 0.9132 | 07m11s |
| 96 BC, Feb 23 | 77 | 0.9229 | 07m19s |
| 91 BC, Oct 22 | 61 | 0.9164 | 08m11s |
| 87 BC, Feb 14 | 58 | 0.9307 | 08m51s |

The longest central duration of an annular eclipse between years -3999 to +6000 (*Five Millennium Catalogue of solar eclipses* https://eclipse.gsfc.nasa.gov/SEcatmax/SE-3999-6000MaxA.html, Eclipse Predictions by Fred Espenak NASA's GSFC), was in 12m23s, which occurred on December 7th 150AD, when the moon was at *Apogee* and the Earth close to *Perihelion* (December 3rd, according to *Starry Night Pro* planetarium software).

In order to detect the starting dates of the Saros Cycle/*Lunar Apokatastasis Cycles* for an extended time span around the Antikythera Mechanism era of 340BC-87BC, Annular Solar eclipses with central duration longer than 07m00s were selected and are presented on Table 1 (the catalogue of solar eclipses by NASA, https://eclipse.gsfc.nasa.gov/SEcat5/SE-0199--0100.html , https://eclipse.gsfc.nasa.gov/SEcat5/SE-0099-0000.html). According to NASA solar eclipse catalogue, only one Saros series - the Saros series 58 - contains very long central duration annular eclipses, i.e. larger than 09m49s, between 200BC and 100BC - the Era of the Antikythera Mechanism construction. Additionally, Saros series 58 contains the second longest central duration annular eclipse between -1999 to +3000, occurred at December 22nd 178BC, https://eclipse.gsfc.nasa.gov/SEsaros/SEsaros058.html. At the same time, the central duration of annular eclipses of the other Saros series, is much lower (08m37s on Saros series 61).

On each of the Saros series 58, the moon returns at its *Apogee* position, therefore it is impossible to occur a total solar eclipse on this Saros series Fig. 3. The annular solar eclipse of December 22nd 178BC presents a very large central duration of 12m8s, which is the largest central duration of any annular eclipse between 2000BC-149AD (notice also the very large width of this eclipse path https://eclipse.gsfc.nasa.gov/SEatlas/SEatlas-1/SEatlas-0179.GIF , relative to the rest ASE, red-colored paths). This eclipse starts just before sunrise-beginning of the day, close to the 1st hour. In this special date December 22nd 178BC, all of the lunar cycles return to their starting position during the start of the day, signifying a *Saros Cycle Apokatastasis*, so this date could be named as "*The Prominent Saros Cycle Apokatastasis*" (defined by the Authors).

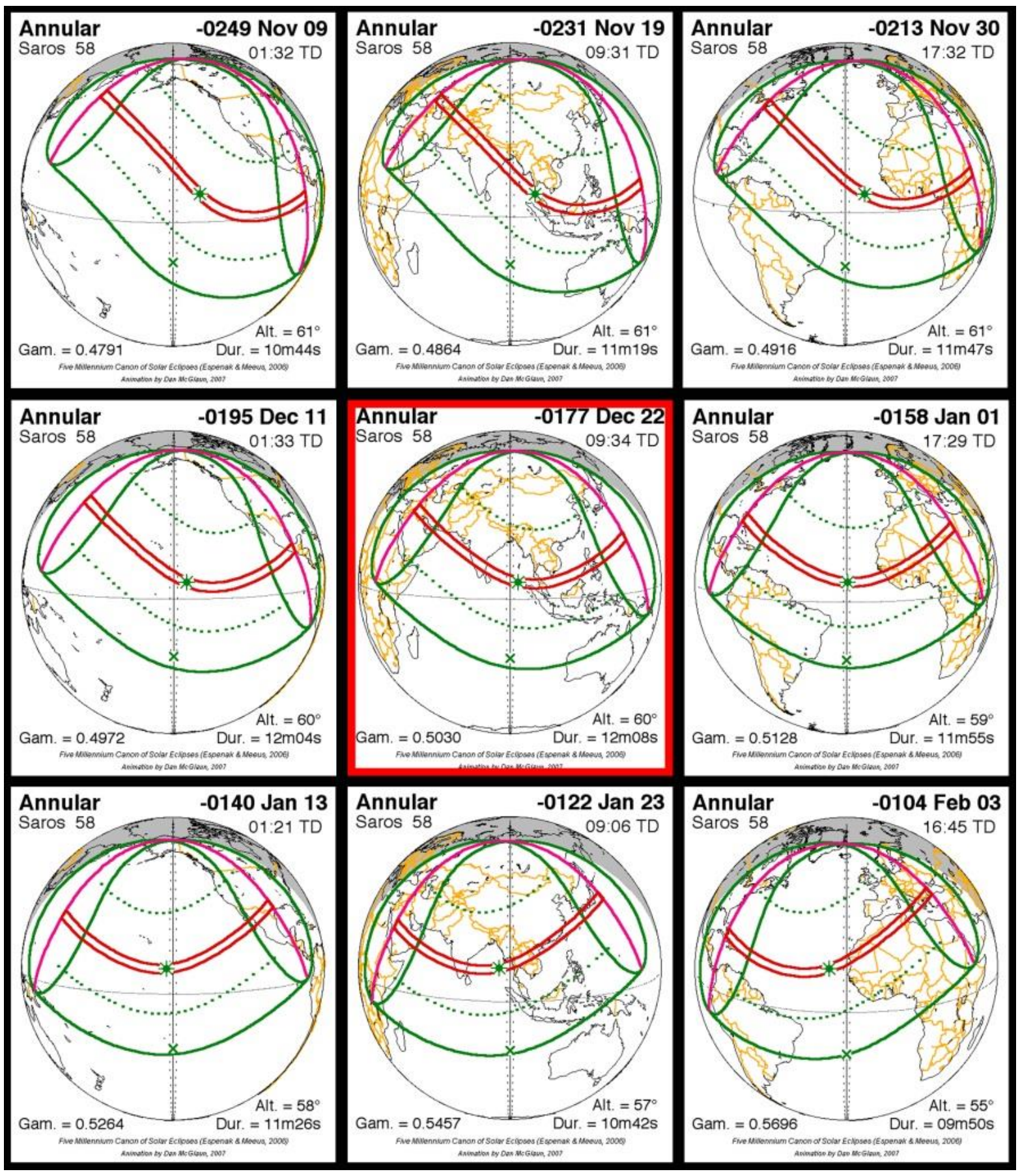

**Fig. 3** Eclipse maps of Saros series 58 (https://eclipse.gsfc.nasa.gov/SEsaros/SEsaros058.html). Saros series 58 starts on 1115BC and closes on 166AD. It consisted of annular solar eclipses in percentage of 61.1%, partial eclipses 38.9% (in dates close to the start/end of the Saros series and only visible very close to the Earth's poles) and 0.0%, total solar eclipses, as a result of the moon's position at Apogee on each Saros repetition. This specific position of the moon during the Apokatastasis of the Saros series 58, presents a particular geometrical symmetry, a large number of annular solar eclipses starting at almost the expected time (by a few minutes difference) and a high repetition on the shadow path positioning on the Earth (concerning the triple Saros), for an extended time around the Antikythera Mechanism era. The eclipses of the middle column start just before sunrise - at about the 1st hour of the day and the eclipses on the last column (after one Saros) start during sunset and are visible from Middle East (and partially visible from Mediterranean, Egypt and Greece). So, two out of three eclipses in one Exeligmos period are visible on the same place because the 8 hour difference is still during the daytime. On-line color image.

Right after the *Prominent Saros Cycle Apokatastasis* date, the Sun crossed into the zodiac sign of Capricorn (see Table 3) on December 23rd 178BC, which marks the beginning of the 3rd WS-Callippic Period (WS CP-3)/calendar of some of the ancient Greek City-States, which had the beginning of the calendar during the Winter Solstice.

In this, very rare coincidence, a WS-Callippic Period begun right after a *Saros Cycle Apokatastasis* and more specifically, right after the *Prominent Saros Cycle Apokatastasis*. The lunar *Apogee* (by measuring the minimum lunar angular velocity, see Chap. 3) and the Solstices dates could be precisely calculated based on naked eye observations, recordings and measurements at the Antikythera Mechanism era (see Geminus calculations on Chap. XVIII, *About Exeligmos*), so this rare coincidence of astronomical events could be also predicted and used by the ancient astronomers.

During the Summer Solstice of June 21th 2020, an annular eclipse occurred https://eclipse.gsfc.nasa.gov/SEgoogle/SEgoogle2001/SE2020Jun21Agoogle.html. The New moon was about on the Ascending Node, but past the Apogee position by 6.2 days before (*Starry Night Pro*; http://www.fourmilab.ch/earthview/pacalc.html).

The *Prominent Saros Cycle Apokatastasis* occurred during the reign of *Ptolemy VI Philometor*, 571 years after the *Era Nabonassar* ("ἔτος ἀπό Ναβονασσάρου", 747B.C., Toomer 1984) and 147 years after the *Great Alexander's death*/Era *Philip III Arrhidaeus* ("ἔτος ἀπό τῆς Ἀλεξάνδρου τελευτής", 323BC), as these two time measuring points, referred in Almagest.

Modern dates (after 1900) in which the Exeligmos/Saros begin are 11th November 1901, 22nd November 1919, 2nd December 1937, 14th December 1955, 24th December 1973, 4th January 1992, 15th January 2010, presenting long duration annular eclipses over 11min, belonging to Saros 141 https://eclipse.gsfc.nasa.gov/SEcatmax/SE-3999-6000MaxA.html.

A particular eclipse occurred on 4th December 2021 http://xjubier.free.fr/en/site_pages/solar_eclipses/TSE_2021_GoogleMapFull.html?Lat=-53.17510&Lng=-51.11367&Elv=-1.0&Zoom=5&LC=1 , http://nicmosis.as.arizona.edu:8000/ECLIPSE_WEB/TSE2021/TSE2021WEB/EFLIGHT2021.html. The New moon (begin of Synodic cycle) was located at Perigee (½ Anomalistic cycle), but faraway from the Node/close to the ecliptic limits (about ¼ of Draconic cycle). Therefore, the totality was only visible from the South pole/Antarctica and also from Scotia and Weddell Sea. The lunar shadow was recorded from the satellite *Deep Space Climate Observatory (DSCVR)* https://earthobservatory.nasa.gov/images/149174/antarctica-eclipsed.

### 3.2 The new numbering of Saros cells, after the Saros Spiral *"Apokatastasis"*

As results by Voulgaris et al., 2021, the preserved Saros parts on Fragment A, are deformed and deviate by their original position. This leads to a correction of the so-far Saros cells initial numbering by Freeth et al. 2008; Carman and Evans 2014; Freeth 2014, Anastasiou et al. 2016; Freeth 2019; Iversen and Jones 2019; Jones 2020 by a factor of -1. This correction based on the geometry and the symmetry of the Antikythera Mechanism design. A very critical observation was the position of the Saros spiral retention bar and the four secret pins, which stabilize the Saros spiral turns on their proper position via the retention bar (Fragment A2), Fig. 4.

The new numbering of the Saros cells also changes the numbering of the preserved eclipse possibility events (also by a correction factor of -1), Table 2. The change, affects the Antikythera Mechanism Epoch date calculated by Freeth 2014; Carman and Evans 2014; Jones 2017; Freeth 2019; Iversen and Jones 2019; Jones 2020).

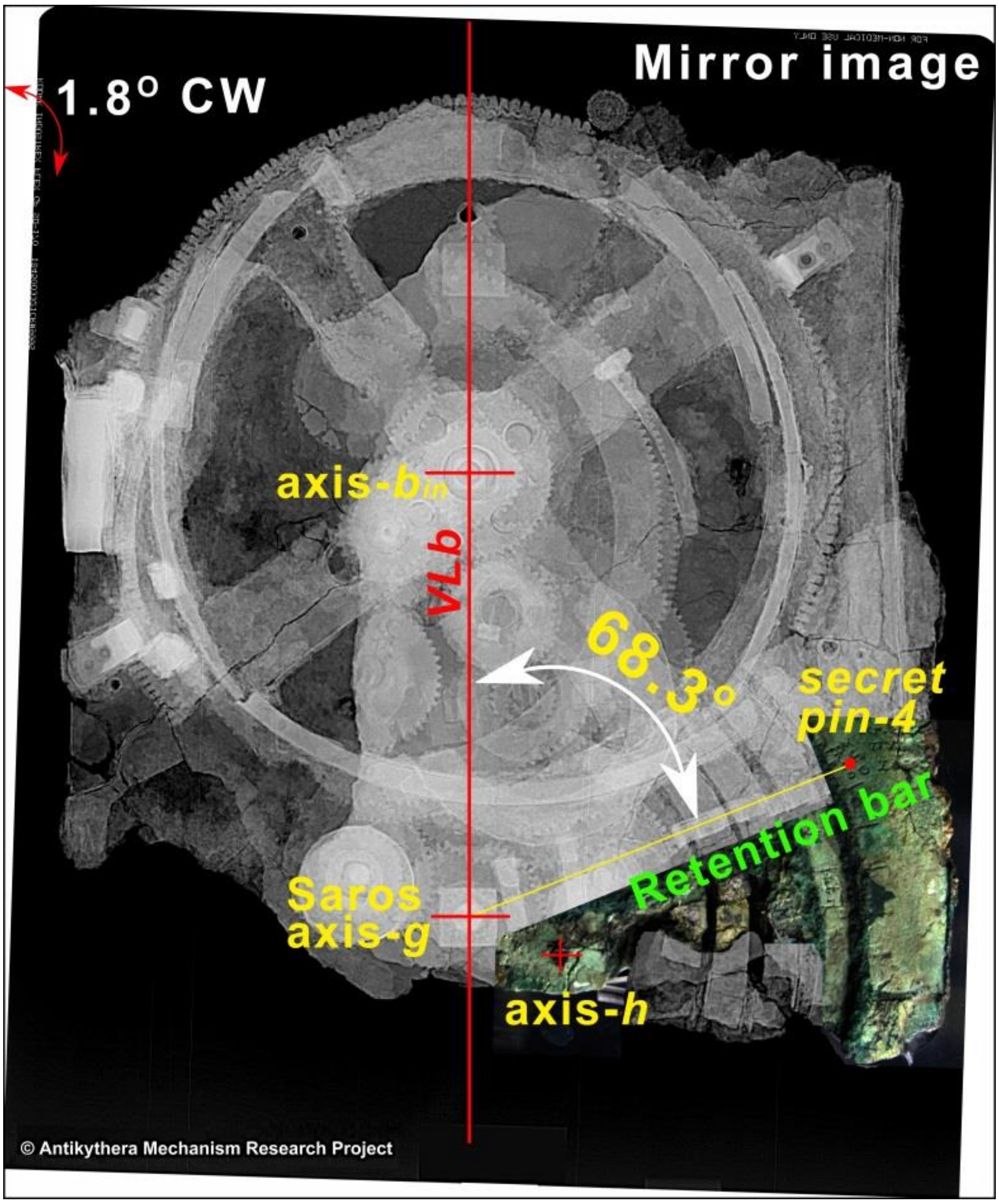

**Fig. 4** AMRP X-ray radiography of Fragment A. The radiography was flipped horizontally and then rotated relative to the axis-$b_{in}$ center by 1.8° CW, so that both the axis-$b_{in}$ and the axis-*g* (Saros pointer) are crossed by the *VLb* mid-perpendicular line. Then, a well aligned and up to scale PTM photograph, was added. The epicenter angle *VLb*/axis-*g*/retention bar pin-4 (red dot), was measured about 68.3°, instead of the ideal value of 60°, when taking into account the Symmetry of the Back plate design. By calculating the corresponding cell position of the retention bar pin-4 as 60°, the corresponding cell is *cell-177* instead of initial numbered *cell-178.* On-line color image.

On *cell-113* (initial *cell-114*) is preserved only one eclipse event, a lunar eclipse (Freeth et al. 2008, Freeth 2014, Anastasiou et al. 2016, Freeth 2019, Iversen and Jones 2019, Jones 2020). A new Sar period (half a Saros cycle) starts from the middle of *cell-113* (Full Moon. Therefore, *cell-1* (one Sar period prior to *cell-113*) should have an (inversed) eclipse event engraved, i.e. a solar eclipse. So, *cell-1* (start of Saros spiral) corresponds to a solar eclipse and should have (at a minimum) an "H" (*ΗΛΙΟΣ*/*HELIOS*) glyph engraved Freeth et al. 2008. Geminus 1880/2002 (VIII, 14) and Plutarch 1936 (Moralia 44,D-E) refer that solar eclipses occurred on the 30th day of the synodic month and lunar eclipses during the15th day of a month.

**Table 2** The eclipse possibility events located on the preserved part of the 1st spiral turn (second column). Initial cell numbering of the eclipse events on the third column, Freeth et al. 2008; Freeth 2014; Anastasiou et al. 2016; Freeth 2019; Iversen and Jones 2019; and Jones 2020, and corrected cell numbering after the Saros parts positional apokatastasis on the fourth column. *Cell-1* (non-preserved) should have engraved either a solar or a solar and a lunar eclipse event.

| **Event index letter** | **Saros spiral Eclipse possibility events** | **Initial cell numbering before the Saros spiral Apokatastasis** (*Freeth et al. 200;, Freeth 2014; Carman and Evans 2014; Anastasiou et al. 2016; Freeth 2019; Iversen and Jones 2019; Jones 2020*) | **New cell numbering, after the Saros spiral Apokatastasis** *(Voulgaris et al., 2021)* |
|---|---|---|---|
| A | *Sun (or Moon and Sun)* | *(cell-2)* | ***Cell-1*** **Saros spiral begin** |
| B | Moon, daytime 2nd hour, Sun, 1st hour | *(cell-8)* | **Cell-7** |
| Γ | Sun, 1st hour | *(cell-13)* | **Cell-12** |
| E | Moon, 6th hour | *(cell-20)* | **Cell-19** |
| Z | Sun, 6th hour | *(cell-25)* | **Cell-24** |
| H | Moon, daytime, 7th hour | *(cell-26)* | **Cell-25** |

## 3.3 The Ancient Greek calendars and the Antikythera Mechanism Parapegma. Correlating the Saros Prominent *Apokatastasis* and the Antikythera Mechanism Front and Back plate Dials

### 3.3.1 Ancient Greek calendars

The ancient Greek calendars were based on the lunar synodic cycle, although they were not precisely synchronized to the tropical year (octaeteris, see Geminus 2002; Evans and Berggren 2006; Samuel 1972). A good approximation between the correlation of the lunar synodic and tropical solar cycles, was firstly the Metonic Cycle ($19^y$).

On ancient times, time keeping and measuring was quite unstable and could be easily lost following a famine or a revolution or a war. Therefore, the correlated astronomical events at the calendar start, offered a specific, unique and rare time reference point, which could be easily detected/re-calculated using astronomical maps, observations and measurements. For example, the selection of a starting date which included at the same time a New moon phase with an equinox or solstice event, defined a characteristic reference-starting point for the starting of the calendar and time measuring (Hannah 2015). Callippus corrected the Metonic Cycle and introduced the Callippic Cycle consisting of 4 Metonic Cycles - 76 years minus 1 day, beginning on (or right after) the New Moon of the Summer Solstice of June 28th 330BC, as also results from Ptolemy's Almagest (Ptolemy 1898; Toomer 1984; Danezis and Theodosiou 1993; Jones 2000).

The ancient Greek City-States used calendars based on the Metonic cycle, but with many differences. Each Greek City-State used different names for the synodic months of the year and often a specific month, was located in a different position on the year, i.e. same month in different season (Iversen 2017; Danezis and Theodosiou 1993; Pritchett 1957). Moreover, the ancient Greek Metonic calendars did not have a common start. For example, the Attic and the Delphi calendar both started on or just after the New moon of the Summer Solstice. The calendar of Macedonia, Lacedaemon (Sparta), Rhodes, Kos, Crete and Miletos started on the New moon of the Autumn Equinox, the Boeotian and Delos calendar started on the New moon of the Winter Solstice and in Chios Island the calendar started on the Vernal Equinox (Thomson 1943, see also different results by Danezis and Theodosiou 1993; Iversen 2017).

**Table 3** The Callippic Period was in use up to the new correction introduced by Hipparchus (ca 100BC). Considering December 25th 330BC as the starting date of the 1st WS-Callippic period, the following 2nd, 3rd and 4th WS-Callippic periods can be calculated by adding 76 years minus 1 day (this day was subtracted in the last 19 year-Metonic segment of each Callippic period)

| Callippic Period/Year | Dates of Callippic Periods beginning on the Winter Solstice (divided in $19^y$) |
|---|---|
| **CP-1** (330-312), year 01-19 | 25 December 330BC |
| CP-1 (311-293) year 20-38 | 25 December 311BC |
| CP-1 (292-274) year 39-57 | 25 December 292BC |
| CP-1 (273-255) year 58-76 | 24 December 273BC |
| **CP-2** (254-236) year 01-19 | 24 December 254BC |
| CP-2 (235-217) year 20-38 | 24 December 235BC |
| CP-2 (216-198) year 39-57 | 24 December 216BC |
| CP-2 (197-179) year 58-76 | 23 December 197BC |
| **CP-3 (**178-160) year 01-19 | 23 December 178BC |
| CP-3 (159-141) year 20-38 | 23 December 159BC |
| CP-3 (140-122) year 39-57 | 23 December 140BC |
| CP-3 (121-102) year 58-76 | 22 December 121BC |
| **CP-4** (102-83) year 01-19 | 22 December 102BC |
| CP-4 (83-64) year 20-38 | 22 December 83BC |

There is a peculiarity associated with the year 330/(329) BC (and also the years before and after these, for several Metonic cycles): The new Moon phases happened during (or close to) the dates of the Summer solstice, Autumn Equinox (September 25th 330BC), Winter Solstice (December 23rd 330BC) and Vernal Equinox (March 21th 329BC) (*Starry Night Pro* software). For those ancient Greek Callippic/(Metonic) calendars that started during the Winter Solstice, the ideal starting date was six synodic months after the Summer Solstice of 330BC, i.e. the Winter Solstice of (24)25 December 330BC, right after the New moon of 23/24 December 330BC (http://www.geoastro.de/planets/moon/pacalc/PeriApCalc.html#patab ; *Starry Night Pro* software). Considering the Winter Solstice (WS) of December 25th 330BC as the starting date of the 1st Callippic Period of an ancient Greek calendar that started at the Winter Solstice (WS-CP1), the dates of the next Metonic Cycles/Callippic Periods, can be calculated (see Table 3).

### 3.3.2 The Parapegma of the Antikythera Mechanism

The Parapegma was located on the Front plate of the Antikythera Mechanism. The Parapegma star events were engraved on the two oblong plates, located on top and bottom position relative to the central Front dial (Bitsakis and Jones 2016a).

On the upper Parapegma plate (PP1), the *column i* (on the left), starts with the seasonal period of Winter: "*Α-ΑΙΓΟΚΕΡΩΣ ΑΡΧΕΤΑΙ ΑΝΑΤΕΛΛΕΙΝ-ΤΡΟΠΑΙ ΧΕΙΜΕΡΙΝΑΙ*" (*1-Capricorn begins to rise - Winter Solstice*). The Summer Solstice events are located on the lower plate (PP2) on the (right) *column iv*: "*Μ-ΚΑΡΚΙΝΟΣ ΑΡΧΕΤΑΙ ΑΝΑΤΕΛΛΕΙΝ-ΤΡΟΠΑΙ ΘΕΡΙΝΑΙ*" (*12-Cancer begins to rise - Summer Solstice*) (Bitsakis and Jones 2016a) Fig. 5.

Two questions naturally arise:

a) If the Metonic calendar of the Mechanism started on the Summer Solstice, what reason prevented the ancient manufacturer to place the Summer Solstice event on the top left of the Mechanism front plate (as the first event)?

b) If the Metonic calendar of the Mechanism started on the Winter Solstice, where that event should be placed on the Front plate Parapegma?

The specific placement of the two Parapegma plates does not affect any mechanical operation of the Mechanism and therefore the ancient manufacturer could easily engrave/adapt the "*Summer Solstice*" event on the top left column, if the Metonic spiral calendar of the Mechanism started during the Summer Solstice (as is valid for the start of the Callippic calendar referred by Ptolemy in Almagest, Goldstein and Bowen 1991). Additionally, the index number of the Summer Solstice, as the 12th event of the second Parapegma Plate (PP2), does not give a primary importance on this event, in contrast to the index number of the Winter Solstice, as the 1st event of PP1. Moreover, the Zodiac Month ring is also free to rotate (as the Egyptian calendar ring, Voulgaris et al. 2018a) therefore the aligning of the Parapegma events relative to the position of the zodiac months is not valid.
It does not seem logical and it's very difficult to justify why the Mechanism Metonic calendar starts during the Summer Solstice and its corresponding Parapegma events are located on the bottom-right position, with the index number 12.

Taking into account that Ancient Greek text is read from *left to right* and from *top to bottom* direction, the discrepancies of the different ancient Greek calendars and according to the specific position of the Winter Solstice event on the Mechanism Parapegma, the Authors believe and consider that the Antikythera Mechanism Metonic calendar, started during a Winter Solstice date. The synodic month on the Antikythera Mechanism started right after a New moon and therefore, the Antikythera Mechanism Metonic spiral start/*cell-1*, begun when the sun and moon were located on (or very close to) the zodiac sign of Capricorn, on a year between 200BC-100BC, considering this time span as the most probable for the construction of the Antikythera Mechanism (Freeth et al. 2006).

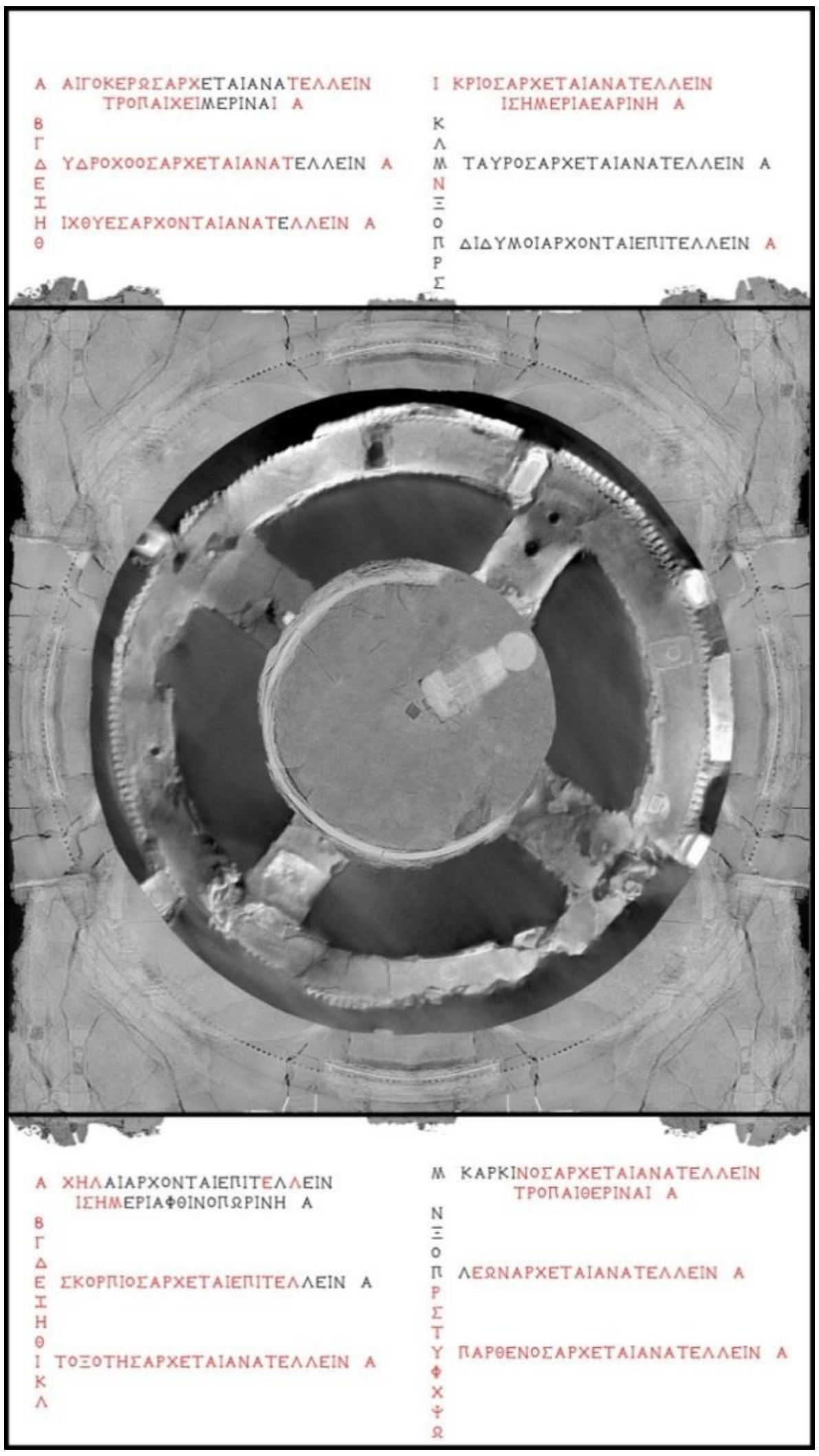

**Fig. 5** A composition of the central Front plate of the Mechanism using multi-combined AMRP radiographies and tomographies of Fragment C, processed by the Authors. The large central hole, the third ring/plate (Voulgaris et al. 2018a) with 365 holes (see also Fig. 6) and the Lunar Disc, are visible. A multilayer AMRP tomography composition of the annual gear *b1,* processed by the authors, is also added. The scheme of the two Parapegma plates PP1 and PP2 were added on top and bottom (Bitsakis and Jones 2016a). The preserved letters/words of the 12 zodiac months and the index numbers (Greek letters) of the Parapegma events are in black letters and the reconstructed/very probable words in red (Bitsakis and Jones 2016a and also some positions according to Authors' opinion). On top left: Capricorn-Winter Solstice, top right: Aries-Vernal Equinox, bottom left: Claws/Libra-Autumn Equinox, bottom right: Cancer-Summer Solstice. On-line color image.

## Results

### 4.1 The positional Apokatastasis of the pointers during the initial calibration date of the Antikythera Mechanism

The unique coincidence of the start of the Winter Solstice-Callippic Period on December 23rd 178BC, right after the "*Prominent Saros Cycle Apokatastasis*" on December 22nd 178BC, could have led the Antikythera Mechanism manufacturer to consider one of these cycles starting date, as the base for the initial calibration date for the pointers' *Apokatastasis* of his creation. Of course, if the Era he lived it was 20 to 40 years after the initial calibration date (see Table 4), he could have chosen as a starting date one or two Metonic/Saros cycles after this specific date.

Considering that the solar eclipses occurred at the last day of each synodic month (30$^{th}$, named *τριακάς-triakas*, see Geminus 2002; Evans and Berggren 2006 on Chap. VIII, §14 and X, §6; Plutarch 1936, Moralia 44D-E), the start of each synodic month is right after the New Moon, as also is valid for the Antikythera Mechanism (*Assembling the Fragment D on the Antikythera Mechanism: Its role and operation*, Authors' submitted work). This means that the synodic month including the *Prominent Saros Cycle Apokatastasis* beginning, started one month before the beginning of the WS-Callippic Period, on 22/23 November 178BC.

One synodic rotation of the Lunar Disc-Input (re-aiming to the Golden sphere-Sun), rotated the Metonic and Saros Pointers by one corresponding measuring unit, that is one cell/synodic month. Therefore, the two pointers during their rotation aimed to the corresponding -same date-cells.

So, the ancient manufacturer had three choices:

a) to define the 1$^{st}$ day of month including the *Prominent Saros Cycle Apokatastasis* as the initial calibration date of the Mechanism on Saros *cell-1*, therefore the Metonic pointer aiming to the last *cell-235* (and after one synodic month to relocate the Metonic pointer on the cell-1) or

b) to define as the initial calibration date of the Mechanism, the date of the WS-Callippic calendar begin on the Metonic *cell-1* and therefore, the Saros pointer aiming to the Saros *cell-2*. This means that the Saros *cell-1* corresponds to the month which includes the start of the *Prominent Saros Cycle Apokatastasis* and should have had (at least) the glyph H (solar eclipse) engraved.

c) Another scenario would be to place the synodic month of the *Prominent Saros Cycle Apokatastasis* on the last *cell-223* (and to relocate the Saros pointer back to the first *cell-1* after one synodic month - full turn of the Lunar Disc-input). However, the last Saros cell cannot be considered as the start of the Saros Cycle, since the pointer's relocation after one synodic turn is not consistent with a well-designed machine, this procedure seems to be awkward and also it is in disagreement to the preserved specific cell position of the eclipse events.

From these three choices, the second one seems to be the more functional and proper in order to avoid the early relocation of the pointers. Therefore, the start of the Metonic spiral corresponds to the starting date/month of the WS-Callippic Period and the start of the Saros spiral corresponds to the starting date/month of the Saros Apokatastasis Cycle. When the Metonic pointer is located on Metonic *cell-1*, the Saros pointer is located on Saros *cell-2*. The differential angle distance of the Metonic and Saros pointers, is the smallest for the dates of November 23rd/December 23 of 178BC (a difference of one cell). Selecting any of the previous or of the next WS-Metonic/Saros Cycles, this distance increases significantly Table 4.

**Table 4** A comparison of the WS-Metonic Cycle starting dates to the Saros Cycle Apokatastasis dates. The correlation of the dates of December 22nd (Saros Cycle) and 23rd (WS-Metonic Cycle) 178BC corresponds to the minimum difference between the Metonic and Saros cells. Before and after this resonance date, the time difference increases rapidly

| WS-Metonic cycle begin cell-1 | Saros Cycle Apokatastasis date | Time difference Synodic month/cell | Angle between Metonic and Saros pointers |
|---|---|---|---|
| 24 December 235 BC | 19 November 232 BC | 36 | ≈236° |
| 24 December 216BC | 30 November 214BC | 24 | ≈155° |
| 23 December 197BC | 10 December 196BC | 12 | ≈77° |
| **23 December 178BC** | **22 December 178BC** | **1** | **≈6.45°** |
| 23 December 159BC | 1 January 159BC | 12 | ≈77° |
| 23 December 140BC | 13 January 141BC | 24 | ≈155° |
| 22 December 121BC | 23 January 123BC | 36 | ≈236° |
| 22 December 102BC | 3 February 105BC | 48 | ≈310° |

The two pointers could have a common start on the corresponding *cell-1* only if each synodic month of the Antikythera Mechanism began exactly on the New moon, but the specific sequence of the preserved Saros events, makes this hypothesis impossible.

The Metonic pointer starts from the internal spiral beginning-first boundary line of *cell-1*, which is the 1st day of *ΦΟΙΝΙΚΑΙΟΣ LA* (Phoinikaios month, Metonic Year-1, Anastasiou et al. 2016a), which corresponds to the 30th day of Poseideon (I)/1st day of Poseideon (II) month, of the Year-1 of the 3rd Callippic Cycle on Ptolemy's Almagest, which started on the Summer Solstice of 330BC (https://webspace.science.uu.nl/~gent0113/astro/almagestephemeris.htm), see Table 5.

The Olympic Games started on July/August 776BC (Perrottet 2004), so 776-178 = 598 = (4X149)+2. So, year 178BC is 2 years after the Olympic Games of 180BC. The Olympic Games started on the 8th or 9th Full moon after the Winter Solstice (Vaughan 2002, Greswell 1862, Thomson 1943) i.e. 8.5 or 9.5 synodic months after December 23rd 180BC, therefore the Games started on August 2nd 180BC (probable date). Every 4 years (≈49.5 synodic months), the Athletic Games pointer completed a full rotation so, 360°/49.5 synodic months =

7.27°/1 synodic month. So, during the initial calibration date of December 23rd 178BC (≈ 29.5 synodic months after the Olympic Games 180BC), the Games pointer should be located 7.27° X 29.5 ≈ 214° after LA-ΟΛΥΜΠΙΑ position ≈ 2X90°+34°, i.e. ≈ 34° after LΓ-ΙΣΘΜΙΑ.

## 4.2 The Front dial pointers position during the initial calibration date

A critical observation was noted by D.S. Price 1974: A deep engraved characteristic straight line on the Front central plate of the Mechanism was made by the ancient manufacturer (Price refers to this line, as Fiducial line), meant to be used as a calibration point/line Fig. 6a,b. After the extended use of the Antikythera Mechanism functional models of the Authors and their attempt to calibrate the models, this line is this line is mandatory for the initial calibration position of the Front dial pointers and scales i.e. the solar and lunar pointer and the two ring scales.

Via a special mechanical procedure, the Golden sphere-Sun pointer aimed to the Front Dial Calibration Line and the pointer of the Lunar Disc-input (Voulgaris et al. 2018b), aimed to the Golden sphere (New moon). The morning of December 23rd 178BC corresponds to the date 18th ΑΘΥΡ (ATHYR) of the Egyptian calendar https://www.staff.science.uu.nl/~gent0113/astro/almagestephemeris.htm , see also Chap. 6.1.

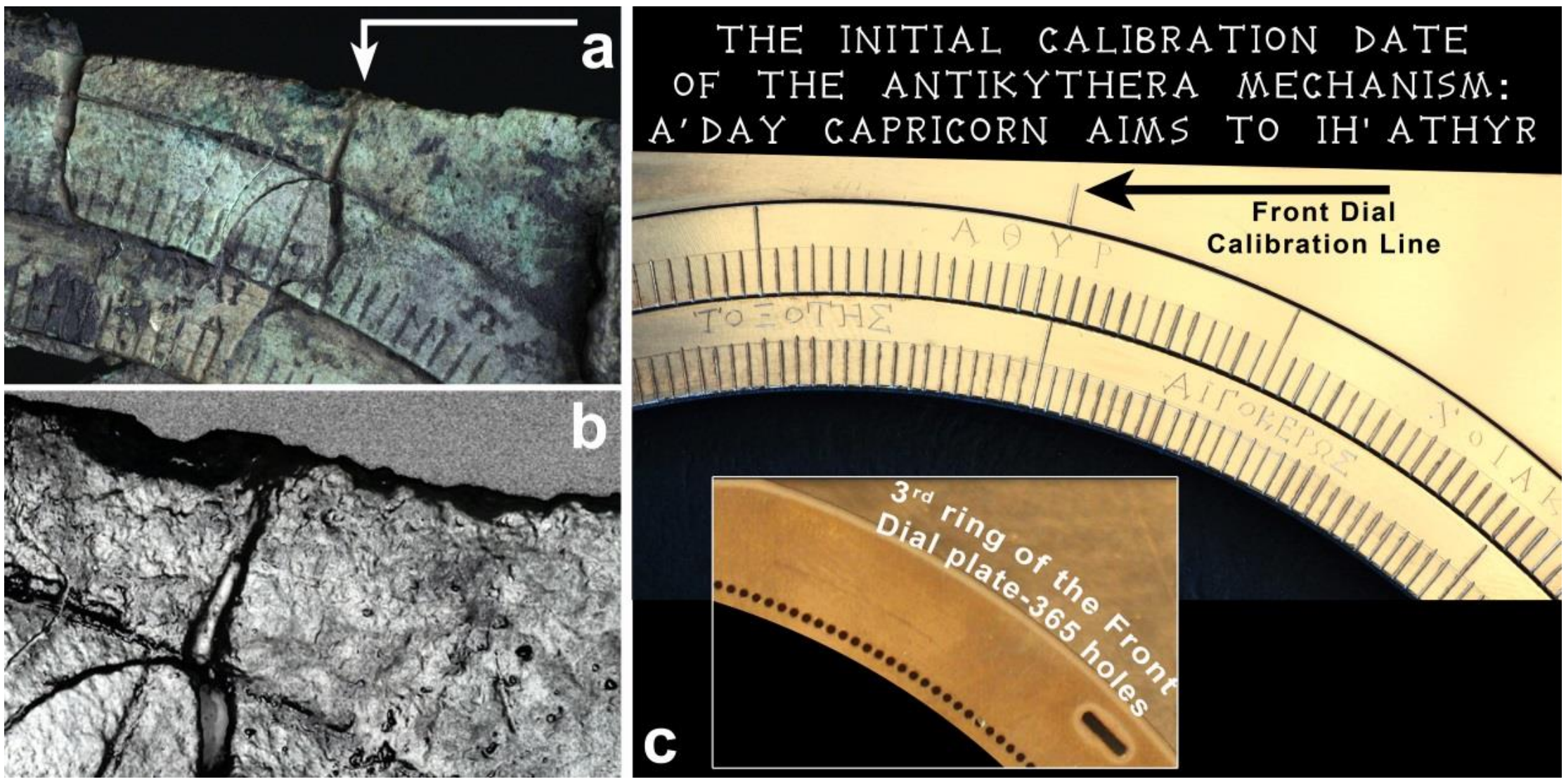

**Fig. 6 (a)** The deeply engraved calibration line on the Front central plate of the Mechanism, is clearly visible on the visual photographs of Fragment C. **(b)** a close up of the calibration line. AMRP PTM images (PTM data courtesy Hewlett-Packard, 2005), processed by the Authors. **(c)** A handmade reconstruction of the Egyptian calendar ring divided in 365 subdivisions/days and the Zodiac month ring, also divided in 365 equal subdivisions/zodiac days, (29 subdivisions-days for the Zodiac months Capricorn and Sagittarius, because of the Earth's perihelion position/fast solar angular velocity, see Tab. 2 of Voulgaris et al. 2018a). The polished bronze (94% Cu, 6%Sn) rings relative position to the Front Dial Calibration Line is presented, according to the initial calibration date of the Antikythera Mechanism. Insert, the unpolished and oxidized (mostly $Cu_2O$-Cuprite, Voulgaris et al. 2019b) bronze 3rd ring with 365 holes, with 0.8mm diameter in a circular distribution. Construction, holes drilling and engraved subdivisions by the first author. Engraved letters by the visual artist C.I. Captured image with a diffused light in perpendicular direction by the first author. On-line color image.

Therefore, the manufacturer rotated the Egyptian calendar ring so that the Golden sphere pointer aimed to the 18th day/subdivision of Athyr. He then rotated the (freely rotatable) Zodiac month ring (Voulgaris et al. 2018a), so that the Golden sphere pointer aimed to the 1st zodiac day/subdivision of ΑΙΓΟΚΕΡΩΣ (Capricorn/Winter Solstice) Fig. 6c. So, the Golden sphere-sun pointer simultaneously aimed to the 18th day of ΑΘΥΡ and to the 1st zodiac day of ΑΙΓΟΚΕΡΩΣ. The lunar phases sphere, which was located on the Lunar Disc-input appeared in its black semi sphere - New Moon phase.

**Table 5** The initial position of the eight pointers of the Mechanism, positioned during the initial calibration date of the Mechanism, December 23rd 178BC

| **Antikythera Mechanism pointer aiming to** | Lunar Disc-input | Lunar Disc pointer | Golden sphere-sun pointer | Golden sphere pointer, 1st Capricorn | Metonic pointer | Saros pointer | Athletic games pointer |
|---|---|---|---|---|---|---|---|
| **Position** | Blackened hemisphere | Golden sphere-sun | 1st zodiac day ΑΙΓΟΚΕΡΩΣ (Capricorn) | 18th day ΑΘΥΡ (Hathyr) Egyptian month | Metonic spiral Cell-1 LA ΦΟΙΝΙΚΑΙΟΣ | Saros spiral Cell-2 | LΓ ΙΣΘΜΙΑ ΠΥΘΙΑ +34° |

## 4.3 A significant observation regarding the calibration of the pin&slot gears

During the positional calibration of the pointers of the Authors' functional models of the Antikythera Mechanism, a very critical parameter was observed, regarding the calibration of gears *k1*/*k2* i.e. the specific position of the *pin* inside the *slot*.

The ingenious *pin&slot* design on *k1*/*k2* gears was implemented by the ancient manufacturer in order to introduce a variable angular velocity on the gearing, representing the variable lunar sky motion (Freeth et al. 2006, see also Chap. 3 of present work). The gears *k2(slot)*/*k1(pin)* are located on board the *e3* gear and one complete rotation of the gears (considering the angle velocity of *e3* gear rotation) corresponds to one Anomalistic cycle.

During the rotation of the *k2*/*k1* gears, the pin travels periodically back and forth inside the slot and therefore the distance between axis *k2* and *pin*, is continuously changing Fig. 7 (Freeth et al. 2006; Voulgaris et al. 2018b). When this distance is largest, the angular velocity of the gear *k1* is the minimum. This position represents the lunar *Apogee* on the Antikythera Mechanism.

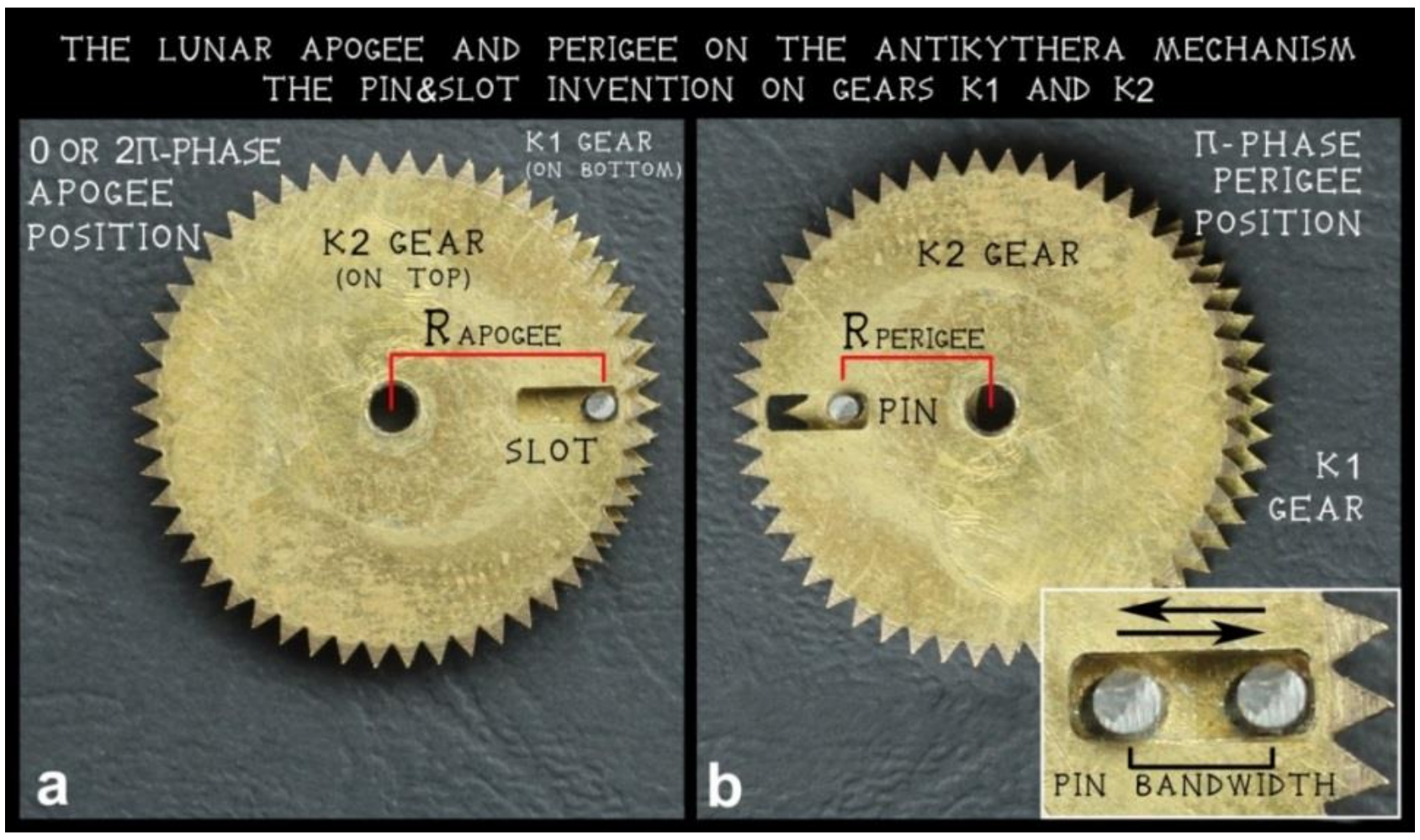

**Fig. 7** The two distinct positions of the pin inside the slot. On insert, a close-up of the pin&slot area, composite image of the two positions. Gears constructed by the first author. On-line color image.

During the assembly and use of the Authors' Antikythera Mechanism functional models, it became evident that there are only two distinct positions on gear system *k1*/*k2*, that are easily and clearly detectable, the "*Apogee*" Fig. 7a and "*Perigee*" Fig. 7b, positions. All of the rest positions of the pin inside the slot do not have any particular geometrical characteristic. Therefore, the specific design of the *pin&slot*, leads to the conclusion that there are only two specific positions for the positional calibration of the *pin* inside the *slot* i.e. specific dates and no other random position/random date, can be used for the *pin&slot* positional calibration.

So, the initial calibration date of the Antikythera Mechanism pointers must be a date corresponding to a lunar *Apogee* or *Perigee*. During the initial calibration date of the Antikythera Mechanism (New moon at *Apogee*), the *pin* of *k1* gear must be placed at the farthest distance from the *k2* gear center Fig. 7a.

## Discussion

## 5.1 The Antikythera Mechanism initial calibration date and the Isia festival

The correlation between the two neighboring dates of 22nd/23rd December 178BC-Winter Solstice, as the initial calibration dates of the Antikythera Mechanism dials, leads to another date coincidence: Based on the Egyptian calendar dating referred by Ptolemy (Ptolemy 1898; Toomer 1984; Goldstein and Bowen 1991; Steele 2000; see also *Almagest Ephemeris Calculator* https://www.staff.science.uu.nl/~gent0113/astro/almagestephemeris.htm), these two dates coincided with the dates of the Egyptian calendar of 17th Athyr 178BC (*Prominent Saros Apokatastasis* beginning) and the 18th Athyr 178BC (beginning of the WS-Metonic/Callippic calendar of the Mechanism). Plutarch refers that the Isia festival was held from 17th to 20th Athyr (Plutarch 1936, Ch.13,C; Ch.39,E; Ch.42,F). The Isia festival started on 17th Athyr with the death of Osiris)/22nd December 178 BC. Geminus, in Chapter VIII, 20

- *About months*, refers that a coincidence occurred about 120 years prior to his present era - the date of the Isia festival coincided with the Winter solstice date (Geminus 1880/2002; Evans and Berggren 2006; Frazer 1914; Jones 1999 and 2017).

The religious festival of Isia, a very important celebration for the ancient Egyptians and Greeks, was dedicated to the gods Isis and Osiris (Plutarch 1874/1889/1936; Richter 2001; Geminus 1880/2002; Evans and Berggren 2006; Frazer 1914; Jones 1999 and 2017).

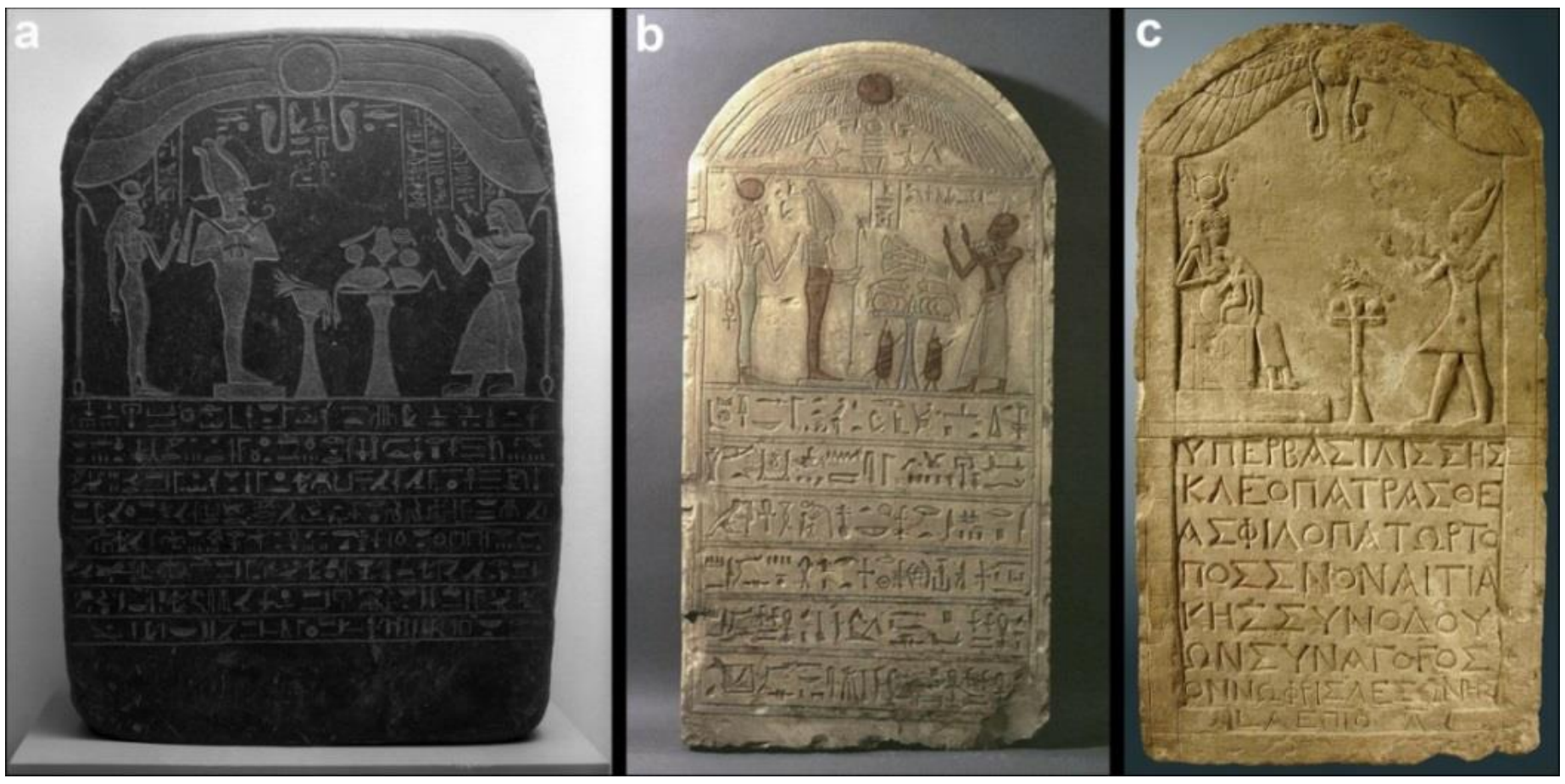

**Fig. 8 (a)** Stela of Irethoreru ca. 775-653 B.C. Beneath the wings of Horus, Irethoreru, (at right), makes an offering to the gods of the underworld, Osiris and his wife, Isis. Brooklyn Museum, USA Gift of Alfred T. White and George C. Brackett, 07.422. **(b)** A stela of 26th Dynasty (?) ca. 664–525 BC. At the top is a winged sun's disk with a legend. Beneath this, Osiris and Isis (on the left) and the stela's owner Nesamun (on the right), Bergmann 1886. Kunsthistorisches Museum Vienna, Austria (09/001, inventory number 119. **(c)** This stele is dated year 1, the 1st day of the month "Epiph" as 2 July 51 BC. The winged disk located in the upper section. Queen Cleopatra (on right) making an offering to the Goddess Isis (on left), who nurses her baby, Horus, Bernand 1992, (inventory number E 27113 RMN, Musée du Louvre, Paris, France). On-line color image.

The Isis and Osiris myth is related with solar and lunar eclipses from ancient time. Plutarch 1936 in *Moralia* 44D-E, describes as an allegorical reference the relation between Isis and Osiris with the Sun, Moon and their eclipses: "*...for the Moon suffers eclipse only when she is full, with the Sun directly opposite to her, and she falls into the shadow of the Earth, as they say Osiris fell into his coffin. Then again, the Moon herself obscures the Sun and causes solar eclipses, always on the thirtieth of the month; however, she does not completely annihilate the Sun, and likewise Isis did not annihilate Typhon.*"

Many ancient Egyptian stones and marble steles depict the Isis and Osiris gods, with a "*winged disk*" symbol at the top. This very old "*winged disk*" symbol was used extensively in Babylon, Assyria, Pharaohs' Egypt and the Ptolemaic Kingdom Fig. 8. Edward W. Maunder (1908) suggested that the "*winged disk*" symbol represented the eclipsed sun and the solar corona. When the solar disk is completely covered by the lunar disc, the solar corona is visible by naked eye Fig. 2 (Pasachoff et al. 2015; Pasachoff et al. 2018; https://www.nasa.gov/feature/how-scientists-used-nasa-data-to-predict-appearance-of-july-2-

eclipse). The bright, elongated and thin coronal streamers appear on the east and west parts of the solar disk, resembling open wings. These coronal formations are (probably) represented on the "winged disks" of the Egyptian steles.
The lunar shadow of the long duration total solar eclipse of August 2, 2027 will cross the land of Egypt, across Luxor (ancient Thebes) and the "*winged sun*" will once again appear in the sky above the Valley of the Kings (https://eclipse.gsfc.nasa.gov/SEgoogle/SEgoogle2001/SE2027Aug02Tgoogle.html , http://xjubier.free.fr/en/site_pages/solar_eclipses/TSE_2027_GoogleMapFull.html).

A date that marks the start of the Isia festival, related to the assassination of Osiris and coinciding with a (visible) solar eclipse at sunrise, as was the 22nd December 178BC, was unique and must have been of very high significance for the Priesthood at that Era. On the other hand, an eclipse happening during the winter solstice was also a rare coincidence and it could be predicted by astronomers of that Era many years before it occurred.

## 6 Conclusions

When the ancient manufacturer designed this unique and remarkable machine, he chose a specific initial calibration date, as a starting position for all of the pointers. The selection of this specific date defined the initial (starting) position for the Mechanism pointers and the specific sequence of the eclipse events was engraved on the Saros spiral cells.
The specific Metonic calendar starting date-Winter Solstice of December 23rd 178BC, right after the *Prominent Saros Cycle Apokatastasis* is an ideal date for the initial calibration of the Mechanism pointers. The Sun and the Moon entered the Tropical of Capricorn (a new WS-Callippic/WS-Metonic Cycle begin), right after the date, New moon was at *Apogee* and at the same time it crossed the Ecliptic, so all of the lunar cycles were on their starting/reset position. The fact that the dates of the Prominent Saros Cycle Apokatastasis and the start of the WS-Callipic cycle coincide with the dates of the important religious Isia festival activities, makes the selection of these dates very enticing.
Of course, the exact date of the construction of the Antikythera Mechanism could be any date after the *Prominent Saros Cycle Apokatastasis*/WS-Metonic/WS-Callippic calendar beginning. Usually, in order to perform time calculations, it is more common to select a date from the recent past rather than one in the future, especially during ancient time, when time calculations and predictions for a large time span, were more uncertain and doubtful than today. This fact could be also the most probable reason for the construction of the Antikythera Mechanism in that era.

## Acknowledgements

*We are very grateful to prof. Xenophon Moussas (National and Kapodistrian University of Athens University) member of AMRP, who provided us with the X-ray Raw Volume data of the Mechanism fragments. Thanks are due to the National Archaeological Museum of Athens, Greece, for permitting us to photograph and study the Antikythera Mechanism fragments. We are very grateful to the Brooklyn Museum, USA, The Kunsthistorisches Museum Vienna, Austria, and the Louvre Museum in Paris, France, for the license and permission to use the photos of the steles. We would like to thank Prof. Zach Ioannou of Sultan Qaboos University of Muscat, Oman, for the hospitality, in order to observe/record via A.Voulg. coronagraphs and spectrographs the Annular Solar Eclipse of December 26th 2019, from Sharqiya Desert and also Prof. Tom Economou of Fermi Institute-University of Chicago, USA, for his help on the preparation of the eclipse expedition and observation. The first and second authors are members of the solar eclipse research team "Totality Hunters", headed by Prof. Jay M. Pasachoff. The team travelled to Chile in order to make white-light and spectroscopic observations of the solar corona during totality from the Cerro Tololo Inter-American Observatory near the Atacama Desert. J.M. Pasachoff passed away in November 2022 and we would like to dedicate this work to his living memory.*